\documentclass[twocolumn,prb,aps,showpacs,groupedaddress]{revtex4}
\usepackage{array}
\usepackage{graphicx}

\def\be{\begin{equation}}
\def\ee{\end{equation}}
\def\bea{\begin{eqnarray}}
\def\eea{\end{eqnarray}}
\def\ba{\begin{array}}
\def\ea{\end{array}}
\def\bdm{\begin{displaymath}}
\def\edm{\end{displaymath}}

\begin{document}

\title{Models of Superconducting Cu:Bi$_2$Se$_3$: \\
single versus two-band description}

\author{S.-K. Yip}

\affiliation{Institute of Physics, Academia Sinica, Taipei, Taiwan}

\date{\today }

\begin{abstract}

Starting from a model Hamiltonian for the normal state of the topological
insulator Bi$_2$Se$_3$, we construct a pseudospin basis for the single-particle
wavefunctions.  Considering weak superconducting pairing near the Fermi surface,
we express the recently proposed superconducting order parameters for
Cu doped Bi$_2$Se$_3$ in this basis.  For the
odd parity states, the $\vec d (\vec k) $-vectors specifying
the order parameter can have unusual
momentum $\vec k$ dependence for certain parameter regimes.  Some peculiar
results in the literature for surface states are discussed
in light of the forms of these $\vec d(\vec k)$'s.  Properties
of the even parity states are also illuminated using this pseudospin basis.
Results from this single-band description are compared with those from
the full two-band model.

\end{abstract}



\maketitle

\section{Introduction}\label{sec:intro}

The recent prediction \cite{FKM07,FK07,HZ09,Xia09}  of the existence of
three-dimensional topological insulators (TI)
 and their experimental confirmation in
Bi$_2$Se$_3$ and related compounds \cite{Hsieh08,Xia09,Hsieh09s,Chen09,Hsieh09n}
 have generated a lot of excitement, as these TI form a new
 class of material which is distinct from ordinary band insulators,
 metals etc, possessing peculiar properties such as topologically
 protected surface states and usual electrodynamics \cite{FK07,Qi08,Qi09}.
 These properties arises from spin-orbit coupling, leading
 to band-inversions at some regions of the Brillouin zone.
 Interestingly, Bi$_2$Se$_3$, when doped with copper, is found
 to become superconducting \cite{Hor10,Wray10,Kriener11}.\cite{BiTe}
   Some unusual properties, such
 as existence of zero bias conductance peak in tunneling experiments
 \cite{Sasaki11,Kirzhner12} and absence of Pauli limiting in
 upper critical field \cite{Bay12}, seem to suggest unconventional
 character of the Cooper pairing, though the situation is not without
 controversy \cite{Levy12}.   There is a lot of attention
 to the theoretical aspects of superconductivity in this compound,
 in particular possible odd parity pairing states.
  Early on, Fu and Berg \cite{FuBerg10}, starting from an
  effective Hamiltonian for the normal state of Bi$_2$Se$_3$ near
  the zero momentum $\Gamma$ point with two orbitals per unit
  cell, considered various models
  of superconducting states with momentum-independent pairs
  when expressed in terms of this basis.  Properties of these
  models have subsequently analyzed by many, mostly
  focusing on the surface states \cite{Hao11,Sasaki11,HsiehFu12,Yamakage12}.

  On the other hand, superconductivity in systems with strong spin-orbit
  coupling have been much studied in the past, in particular
  in the context of heavy fermions
  \cite{Anderson84,VG84,Ueda85,Blount85,Sigrist91,Yip93,Joynt02}.  There, the usual language used is
  that the normal quasiparticles are described by a pseudospin basis,
  obeying certain symmetry properties, and then the superconducting
  state order parameter and pair wavefunctions are expressed in terms of this basis.
  The superconducting states then can be classified by crystal symmetries
  into group representations, and the pairing states are
  expressed in terms the sets of basis functions appropriate
  to the relevant group representations.
  In this formulation, for crystals with inversion symmetry,
  even parity superconductors always have pairing
  wavefunctions even in momentum $\vec k$ and are singlets in
  pseudospins, whereas odd parity superconductors have pairing
  wavefunctions that matrices in pseudopsin space with each
  opponent odd in $\vec k$.  This matrix structure is usually
   expressed in terms of a "d-vector" which specifies the
   corresponding pseudospin structure via familiar Pauli $2 \times 2$ matrices.
   Properties of the superconductors can then be directly obtained
   by examining these order parameters, including the possibility of
   surface states.

   Many questions then arise.  If the chemical potential $\mu$,
    of the doped
   Bi$_2$Se$_3$  is sufficiently large compared with the pairing potential $\Delta$,
   (which is likely to be the case since $\mu \sim 0.4 eV$ according to \cite{Wray10},
   whereas the transition temperature is $< 4$K.  The measured gap
   in tunneling is indeed $ \sim meV$, though the precise value is
   controversial \cite{Sasaki11,Levy12}), pairing should effectively
   only take place within one normal state band.  How well then can we understand
   the superconducting properties of Cu:Bi$_2$Be$_3$ within a single band
   picture?  What is the pairing order parameter in the pseudospin basis?
   Can we understand the Andreev bound states in this
   way?  In particular, what is the origin of the very peculiar dispersions
   found for the odd parity states found in \cite{Hao11,Sasaki11,HsiehFu12,Yamakage12}?

   In this paper, we report such an attempt.  In Sec \ref{sec:back},
   we review the model of \cite{FuBerg10}.  Pseudospin wavefunctions
   are constructed in Sec \ref{sec:wf}.  They would be applied to
   the superconducting phases, first for the bulk in \ref{sec:bulk},
   then the surface states in \ref{sec:surface}.  We shall show that many
   of the results in the literature can be understood in this way.
   We conclude in \ref{sec:concl}. The Appendix gives further discussions
   on the surface states and topology of some of the superconducting phases.

   \section{Model} \label{sec:back}

  In this section, we review the model of \cite{FuBerg10}.  We begin with
  the normal  state.  The effective Hamiltonian which captures the physics
  near $\vec k = 0$ is given by
  \be
  H_N (\vec k) = m \sigma_x + v_z k_z \sigma_y + v \sigma_z (k_x s_y - k_y s_x)
  \label{HN}
  \ee
  Here $\vec k$, $k_x, k_y, k_z$ represent the wavevector and its components,
  $v_z$, $v$ are velocities, $\sigma_z = \pm 1$ represents the two
  (mainly) $p_z$ orbitals in the quintuple layer of Bi$_2$Se$_3$,
  and $s_{x,y,z}$ are the Pauli matrices for the spin.
  Equation (\ref{HN}) is basically a Dirac Hamiltonian.
   The energies for the quasiparticles are $E_{\vec k} = \pm \epsilon_{\vec k}$
   where $\epsilon_{\vec k} = ( m^2 + v_z^2 k_z^2 + v^2 k_\|^2)^{1/2}$
    with $k_\| \equiv \sqrt{k_x^2 + k_y^2}$,
   thus consist of a conduction and a valence band.
   Bi$_2$Se$_3$ possesses $D_{3d}^5$ ($R\overline{3}m$) symmetry,
   which includes parity. Eq (\ref{HN}) is indeed
  invariant under the parity when this operator is taken as $\sigma_x$ \cite{FuBerg10}:
   Eq (\ref{HN}) is left unmodified if we substitute
  $\sigma_{y,z} \to - \sigma_{y,z}$, $ \vec k \to - \vec k$, with
  all other variables unaltered.  Equation (\ref{HN}) is actually
  invariant under the higher symmetry group $D_{\infty h}$
  as it is obviously unchanged under any continuous rotation
  about the $z$-axis instead of only $2 \pi/3$ for $D_{3d}$.
   One can check easily that (\ref{HN}) is
  invariant under reflection about any vertical reflection planes.
  Since it is rotationally symmetric about $z$, it is sufficient
  to check any one single vertical reflection plane.
  For example, under reflection in the x-z plane,
  $k_y \to - k_y$, $s_{x,z} \to - s_{x,z}$, (\ref{HN}) remains indeed
  unchanged. \cite{compare}

  To discuss the surface states associated with the TI, boundary conditions
  for the wavefunction are needed.  Unfortunately, this point seems
  to be somewhat controversial \cite{HsiehFu12,Liu10,ZKM12}.
  For definiteness, we follow \cite{FuBerg10,HsiehFu12} here.
   The boundary condition for the wavefunction $|\Psi>$
   at the $z=0$ plane for a crystal occupying
  $z < 0$ is taken to be $\sigma_z |\Psi> = |\Psi>$,
  and hence it has no projection in the $\sigma_z = -1$ orbital.
  Topological surface states in the form of a Dirac cone exist when  $\rm{sgn} (m v_z) < 0$.\cite{HsiehFu12}
  For spin along $\hat z \times \vec k$, the bound state energy is
  $E_b = v k_\|$.  Hence, the positive energy branch has spin along
  $v \hat z \times \vec k$.  To account for the
  situation of Be$_2$Si$_3$, \cite{Hsieh09n} we need to
  take $v < 0$, though we shall consider arbitrary relative signs of $v$,
  $m$ and $v_z$ below for comparison purposes.  Since
  the bulk energies are given by $\pm ( m^2 + v_z^2 k_z^2 + v^2 k_\|^2)^{1/2}$, and so within this model,
  the surface states are always separated from the continuum for any given $k_\|$.

  Now we consider the superconducting states, first for the bulk.
   For time-reversal and inversion symmetric
  systems, there is a pair of degenerate states at any given momentum $\vec k$,
  forming a pseudospin $1/2$.  Cooper pairing occurs between opposite momenta
  $\vec k$ and $- \vec k$, and can be classified into even parity,
  pseudospin singlet and odd parity, pseudospin triplet states.\cite{Anderson84}  These
  superconducting states
  can further be classified by their different symmetries under the
  crystal symmetries into different representations in group theory.\cite{VG84,Ueda85,Blount85}
  These representations depend only on the point group (but  not the space group).
  Possible forms of the corresponding momentum and pseudospin dependence in each
   group representation expressed in the form of basis functions:
   the general form of the order parameter can be a linear combination
   of the independent basis functions of the same symmetry, each
   term possibly multiplied by a momentum dependent function which
   is invariant under the particular group under consideration.
   They have in particular been listed for the cubic $O_h$, tetragonal $D_{4h}$, and hexagonal
   $D_{6h}$ groups \cite{VG84,Ueda85,Blount85,Sigrist91,Yip93,Joynt02}.
   For $D_{6h}$ we have
   the group representations $A_1$, $B_1$, $A_2$, $B_2$, $E_1$ and $E_2$,
   with each of the above either even ($g$) or odd ($u$) parity.
    The corresponding table for $D_{3d}$ appropriate for Bi$_2$Se$_3$
   was not listed in these references, but can be trivially obtained from those
   for $D_{6h}$ since $D_{6h}$ would reduce to $D_{3d}$ if we discard
   rotations about $\hat z$ of odd multiples of $2 \pi/6$ and three of
   the horizontal rotational axes. In this case, $A_1$ is no longer distinguishable
   from $B_1$, and similarly for $A_2$ and $B_2$, and $E_1$ and $E_2$.
   The resulting group representations and their basis function are
   listed in the first two columns of table \ref{table}, following Ref \cite{Yip93}.
   For simplicity, we do not list all the possible independent
   basis functions, but mainly those which would appear again in the
   later part of this paper.  For the complete basis function set, we refer the readers to
   the literature \cite{Yip93,Joynt02}.

   In \cite{FuBerg10}, various types of momentum independent (local) pairing in
   the orbital and spin basis of eq (\ref{HN}) were considered.  Since this formulation
   involves states of eq (\ref{HN}) without {\it a priori} distinguishing the conduction
   and valence band, we shall refer to this as the "full two-band description" \cite{twoband}
     The symmetry
   of these states were already discussed in \cite{FuBerg10}, and we list
   each these states with their corresponding symmetries in column (iii)
   of Table \ref{table}.  In this table, we have followed the notation of \cite{FuBerg10}
   and use $1,2$ to label the two orbitals instead of $\sigma_z = 1, -1$ of eq (\ref{HN}).
   For convenience of comparison with other
   works in the literature \cite{Hao11,Sasaki11,Yamakage12}, we also list
   in columns (iv) and (v) these order parameters in matrix form,
   which we shall denote $\bf \Delta^I$ and $\bf \Delta^{II}$ and
   referred to as "Nambu I" and "Nambu II".   $\Delta^I$ is the
   matrix order parameter
   in the ordinary Nambu notation after generalization
   to two orbitals, that is, if we use the operators as
   $(c_{\sigma \uparrow}, c_{\sigma, \downarrow}, c^{\dagger}_{\sigma \uparrow},
   c^{\dagger}_{\sigma, \downarrow})$, where $c_{\sigma s}$ and $c^{\dagger}_{\sigma s}$'s are annihilation
   and creation operators, and $\sigma = \pm 1$ the two orbitals.  If we use instead
   $(c_{\sigma \uparrow}, c_{\sigma, \downarrow},
   c^{\dagger}_{\sigma, \downarrow}, -c^{\dagger}_{\sigma \uparrow})$, as
   done in \cite{HsiehFu12},
   (or  $(c_{\sigma \uparrow}, c_{\sigma, \downarrow},
   -c^{\dagger}_{\sigma, \downarrow}, c^{\dagger}_{\sigma \uparrow})$, as
   in \cite{Yamakage12})
     the order parameter matrix is $\bf \Delta^{II}$.
   The two notations are related simply by ${\bf \Delta^I}  = {\bf \Delta^{II}} ( i s_y)$
   ( ${\bf \Delta^I}  = {\bf \Delta^{II}} (- i s_y)$ ).
   The factorizing out of $(i s_y)$ to the right has the effect of what has
   been done in the $^3$He literature \cite{Leggett75}, where the order parameter matrix
    is written as $ (\vec d \cdot \vec s ) ( i s_y)$, so
   that $\vec d$ transforms as a vector under spin-rotations.  In this
   way, it is clear from column (v) that the two entries listed under $E_{u}$ are related by a
  $\pi/2$ rotations about the $z$-axis.  In making this table, we have
  made use of the gauge symmetry of superconductivity to simplify the matrices
   (by removing factors like $\pm 1$ or $\pm i$).  However, we have
   kept the correct relative phase between the two partners within $E_u$,
   so that they have the correct relative transformation properties.

  \begin{widetext}

    \begin{table}

\begin{tabular}{@{}llllll@{}}
\toprule
(i)  & & (ii) & (iii) & (iv) & (v) \\
\colrule
even parity & & & & &\\
\colrule
$A_{1g}$ & &1 & $|1 \uparrow 1\downarrow> + | 2\uparrow 2\downarrow>$ & $i s_y$ &  1 \\
           & &   & $|1 \uparrow 2\downarrow> - |  1\downarrow 2\uparrow> $ &
             $ i \sigma_x s_y $ & $\sigma_x$ \\
$A_{2g}$ & & {\rm Im} $k_+^6$ & & & \\
$E_{g} $ & & {\rm Re} $k_z k_+$ & & &  \\
 & & {\rm Im} &  & &  \\
\colrule
odd parity & &  & & & \\
\colrule
$A_{1u}$ & & $k_z \hat z$; $k_x \hat x + k_y \hat y$ &
    $|1 \uparrow 2\downarrow> + |  1\downarrow 2\uparrow> $ & $\sigma_y s_x$ & $\sigma_y s_z$ \\
$A_{2u}$ & & $k_x \hat y - k_y \hat x$ &
   $|1 \uparrow 1\downarrow> - | 2\uparrow 2\downarrow>$ & $\sigma_z s_y$ &  $\sigma_z$\\
$E_{u}$ & & {\rm Re} $ k_{+} \hat z$; $k_z \hat r_{+}$; $k_{+}^2 k_z \hat r_{-}$ &
       $ i ( | 1 \uparrow 2 \uparrow > - | 1 \downarrow 2 \downarrow>) $ & $- \sigma_y s_z$ & $\sigma_y s_x$ \\
       & & {\rm Im}   &
       $  | 1 \uparrow 2 \uparrow > + | 1 \downarrow 2 \downarrow> $ & $i \sigma_y$ & $\sigma_y s_y$ \\
  \botrule
  \end{tabular}
  \caption{Representations (column (i)), basis functions (column (ii)) and pairing wavefunctions
  (column (iii)) for the superconducting phases considered in this paper \cite{others}.
   The matrix form of the order parameters
  for (iii) are given in (iv) and (v) for the Nambu-I and Nambu-II representations.
    Here $\hat r_{\pm} = \hat x \pm  i \hat y$, $k_{\pm} =
  k_x \pm i k_y$. } \label{table}
  \end{table}

\end{widetext}

 Table \ref{table} shows the correspondence between each local pair
   and the possible basis functions.  However, it does not tell
   us directly what exactly the momentum dependences are, since
   multiplication of any basis function by a function invariant
   under the crystal symmetry is also an as good basis function.
   It also does not tell us, in particular for the case of odd-parity
   pairing, what linear combinations between the different inequivalent
   basis functions (e.g. $k_x \hat x + k_y \hat y$ and $k_z \hat z$ for $A_{1u}$)
   that we should take. We shall see later that this information is
   in fact useful for understanding the surface state spectrum.
   To do this, we have to first construct
   the single particle pseudospin wavefunctions obeying the correct
   crystal symmetries, as we shall do in Section \ref{sec:wf}. We
   shall then take up the task of showing how the pairings listed in column (iii) correspond
   to the basis functions listed in column (ii) in Sec \ref{sec:bulk}.

   \section{Pseudospin Wavefunctions} \label{sec:wf}

   Now we construct the pseudospin wavefunctions.  For each $\vec k$, we shall
   denote the two degenerate states by $|\vec k,\alpha>$ and $|\vec k,\beta>$.
   We shall demand that the corresponding pairs at $- \vec k$ are related
   to those at $\vec k$ by the relations
   \bea
   | - \vec k, \alpha > &=& P | \vec k, \alpha > \label{P} \\
   | - \vec k, \beta > &=& T |\vec k, \alpha > \label{T}
   \eea
   where $P$ denotes the parity and $T$ the time-reversal.
   $P$ was already taken to be $\sigma_x$, and
   we shall take $T$ as $ (- i s_y)$ times the complex conjugate.
   Correspondingly, we have
   \be
   | \vec k, \beta > = P T |\vec k, \alpha > \label{PT}
   \ee
   and
   \be
   | - \vec k, \beta > = P |\vec k, \beta > \label{P2}
   \ee
   Note that $T |\vec k, \beta> = - | - \vec k, \alpha > $, as $T^2 = -1$.
   $|\vec k, \alpha >$ and $|\vec k, \beta>$ are also required to
   satisfy certain rotational symmetry properties, to be treated in details below.
   These requirements basically enable us to roughly think of
   $\alpha$ and $\beta$ as ``spin-up" and ``spin-down" respectively.\cite{otherbasis}
   We shall first deal with eq (\ref{P}) and (\ref{T}), ignoring
   the rotational properties for the moment.  We shall thus first find an intermediate
   basis $|\vec k, \alpha'>$ and $|\vec k, \beta'>$ obeying
   equations (\ref{P}) and (\ref{T}) without worrying about the rotational
   properties.

   For this purpose, let us first introduce spin wavefunctions
   which diagonalize the spin part of eq (\ref{HN}), using
   \be
   |\hat s = \hat z \times \hat k > = \frac{1}{\sqrt{2}}
   \left( \ba{c}
   1 \\
   i e^{i \phi_{\vec k}} \ea \right)
   \ee
   \be
   |\hat s = -\hat z \times \hat k > = \frac{1}{\sqrt{2}}
   \left( \ba{c}
   i e^{-i \phi_{\vec k}} \\ 1 \ea \right)
   \ee
   for spins along $\pm \hat z \times \vec k$. Here
   $\phi_{\vec k}$ is the azimuthal angle of $\vec k$
   in the $x-y$ plane.  These spin wavefunctions satisfy
   $ (k_x s_y - k_y s_x) |\hat s = \pm \hat z \times \hat k>
    = \pm k_\| |\hat s = \pm \hat z \times \hat k> $.
    It is then straight-forward to diagonalize eq (\ref{HN}).
    We shall define $|\vec k, \alpha'>$ for $k_z > 0$
    (the ``northern hemisphere") to be the one
    associated with spin along $\hat z \times \hat k$:
    \be
    |\vec k, \alpha'> \equiv \frac{1}{\sqrt{2} \mathfrak{N}}
    e^{ i \vec k \cdot \vec r}
    \left( \ba{c} E_{\vec k} + v k_\| \\
     m + i v_z k_z \ea \right)
    \left( \ba{c}
   1 \\
   i e^{i \phi_{\vec k}} \ea \right)
   \label{ka1}
   \ee
   where the first column matrix denotes the part in orbital space
   and the second part denotes the spin space. Here $E_{\vec k}$
   is the energy of the particle (which can be $\pm \epsilon_{\vec k}$), and
    $\mathfrak{N} \equiv [ 2 E_{\vec k} ( E_{\vec k} + v k_\|)]^{1/2}$
   is a renormalization factor.
   The other state $|\vec k, \beta'>$ for $k_z > 0$, as
   well as states for $k_z < 0$, are obtained by the symmetry
   requirements (\ref{P}),(\ref{T}) and (\ref{PT}) \cite{notekz0}.
   We thus have, for $k_z > 0$,
    \be
    |\vec k, \beta'> \equiv \frac{1}{\sqrt{2} \mathfrak{N}}
    e^{ i \vec k \cdot \vec r}
    \left( \ba{c} m - i v_z k_z  \\
      E_{\vec k} + v k_\| \ea \right)
    \left( \ba{c}
   i e^{ - i \phi_{\vec k}} \\
    1 \ea \right)
    \label{kb1}
   \ee
   and
     \be
    |- \vec k, \alpha'> \equiv \frac{1}{\sqrt{2} \mathfrak{N}}
    e^{ - i \vec k \cdot \vec r}
    \left( \ba{c}m + i v_z k_z  \\
      E_{\vec k} + v k_\| \ea \right)
    \left( \ba{c}
   1 \\
   - i e^{i \phi_{- \vec k}} \ea \right)
   \label{-ka1}
   \ee
       \be
    |- \vec k, \beta'> \equiv \frac{1}{\sqrt{2} \mathfrak{N}}
    e^{ - i \vec k \cdot \vec r}
    \left( \ba{c} E_{\vec k} + v k_\|  \\
      m - i v_z k_z \ea \right)
    \left( \ba{c}
   - i e^{- i \phi_{-\vec k}} \\
    1 \ea \right)
    \label{-kb1}
   \ee
   We have written the last two wavefunctions for wavevectors in
   the ``southern hemisphere"
   using the labels $ - \vec k$ with $k_z > 0$.  This is for convenience
   later since we shall always be consider Cooper pairs between $\vec k$ and
   $- \vec k$, and it is sufficient to write these pairs with $k_z > 0$ for the pair
   due to the fermionic antisymmetry of wavefunctions.
   Note that $\phi_{- \vec k} = \pi + \phi_{\vec k}$.

   We now proceed to find the wavefunctions $| \vec k, \alpha > $ and
   $|\vec k, \beta>$ with the desired $P$, $T$, and rotational properties.
   One can of course directly study the wavefunction themselves.  However,
   a more convenient way to proceed is to evaluate some physical quantity
   with known transformation properties ({\it c.f.} \cite{Blount85}).
   For this, we consider the spin operators projected onto our
   two-dimensional Hilbert space $| \vec k, \alpha > $ and $|\vec k, \beta>$
   for each $\vec k$ point. These operators are thus then
   also $2 \times 2$ matrices.  This spin operator is related to the effective
   magnetic moment of our quasiparticles, if the orbital contributions
   can be ignored (which can indeed be the case if the relevant orbitals
   are just $p_z$ and the mixing to $p_x \pm i p_y$ can be ignored).
   We shall therefore denote them as $\vec{\mathfrak{m}}^{\rm eff}$.
   Anyway,  the operators for this effective spin moment $\mathfrak{m}^{\rm eff}_j$,
   $j = x, y, z$, are simply
   \be
   \mathfrak{m}_{j}^{' \rm eff} (\vec k) =
   \left( \ba{cc} < \vec k, \alpha'| s_{j} | \vec k, \alpha'> &
      < \vec k, \alpha'| s_{j} | \vec k, \beta'> \\
      < \vec k, \beta'| s_{j} | \vec k, \alpha'> &
      < \vec k, \beta'| s_{j} | \vec k, \beta'> \ea
      \right)
   \label{def-seff}
   \ee
   where the $s_{j}$ inside the matrices are the spin Pauli matrices
   as in eq (\ref{HN}).  The prime $'$ is to remind us that we are
   using the $| \vec k, \alpha' > $ and $|\vec k, \beta' >$ basis at this
   moment.  Viewed as operators, we thus
   have
   \bdm
       \mathfrak{m}_{j}^{' \rm eff} (\vec k) =
   \sum_{\gamma, \gamma'= \alpha', \beta'} | \vec k, \gamma >
   < \vec k, \gamma | s_j | \vec k, \gamma'> < \vec k, \gamma'|  \
   \edm
   Straight-forward calculations using eqns (\ref{ka1}) and (\ref{kb1})
   give, for $k_z > 0$, (those with $k_z < 0$ can be found later by
   using eq (\ref{P}) and (\ref{T}), this guarantees the correct
   properties under parity and time-reversal)\cite{notePT}
   \be
      \mathfrak{m}_{x}^{' \rm eff} (\vec k) =
      \left(
      \ba{cc} - {\rm sin} \phi_{\vec k} &
      |A_{\vec k}| {\rm cos} \phi_{\vec k} e ^{ - i (\phi_{\vec k} + \alpha_{\vec k})} \\
      |A_{\vec k}| {\rm cos} \phi_{\vec k} e ^{  i (\phi_{\vec k} + \alpha_{\vec k})} &
      {\rm sin} \phi_{\vec k}  \ea \right)
      \label{sx1}
   \ee

   \be
      \mathfrak{m}_{y}^{' \rm eff} (\vec k) =
      \left(
      \ba{cc}  {\rm cos} \phi_{\vec k} &
      |A_{\vec k}| {\rm sin} \phi_{\vec k} e ^{ - i (\phi_{\vec k} + \alpha_{\vec k})} \\
      |A_{\vec k}| {\rm sin} \phi_{\vec k} e ^{  i (\phi_{\vec k} + \alpha_{\vec k})} &
       - {\rm cos} \phi_{\vec k}  \ea \right)  \ ,
      \label{sy1}
   \ee
   and
   \be
      \mathfrak{m}_{z}^{' \rm eff} (\vec k) = |A_{\vec k}|
      \left(
      \ba{cc}  0  &
       i e ^{ - i (\phi_{\vec k} + \alpha_{\vec k})} \\
       - i e ^{  i (\phi_{\vec k} + \alpha_{\vec k})} & 0 \ea \right)
      \label{sz1}
   \ee
   where
   \be
   A_{\vec k} \equiv |A_{\vec k}| e^{ - i \alpha_{\vec k}}
   \equiv \frac{2}{\mathfrak{N}^2}
   ( E_{\vec k} + v k_{\|} ) ( m - i v_z k_z)
   \ee
   is a factor generated by the overlap of the orbital wavefunctions
   in eq (\ref{ka1}) and (\ref{kb1}), and so
   \be
   |A_{\vec k}| = \frac{ ( m^2 + v_z^2 k_z^2 )^{1/2}}{|E_{\vec k}|}
   \label{Aabs}
   \ee
   and
   \be
   e^{ - i \alpha_{\vec k}} = ({\rm sgn} E_{\vec k})
      \frac{ m - i v_z k_z } {( m^2 + v_z^2 k_z^2 )^{1/2}}
      \label{alpha}
      \ee
    The set of $2 \times 2$ matrices $\rho'_{1,2,3}$, \cite{123}
    with $\rho'_1 (\vec k) \equiv | \vec k, \alpha'> < \vec k, \beta'| +
      | \vec k, \beta'> < \vec k, \alpha'|$ etc do not
      yet have the desired transformation properties
      under rotation.
      We now construct a new basis $|\vec k, \alpha>$, $|\vec k, \beta>$
      so that the corresponding Pauli matrices
      $\rho_{x,y,z}$ for the pseudospin do
      transform like an axial vector.  Since the system has
      complete rotational symmetries about $\hat z$, we must
      require $\mathfrak{m}_z^{\rm eff} \propto \rho_z$.
      To do this, we simply have to find a basis
      so that  $\mathfrak{m}_z^{\rm eff} $ is diagonalized.
      This can be done by choosing
      \be
      |\vec k, \alpha> = \frac{e^{i \theta_{\vec k}}}{\sqrt{2}}
      \left( |\vec k, \alpha'> - i e^{ i (\phi_{\vec k} + \alpha_{\vec k})}
        |\vec k, \beta'> \right)
      \label{transf1a}
      \ee
      and
   \be
      |\vec k, \beta> = \frac{e^{-i \theta_{\vec k}}}{\sqrt{2}}
      \left( |\vec k, \beta'> - i e^{ -i (\phi_{\vec k} + \alpha_{\vec k})}
        |\vec k, \alpha'> \right)
      \label{transf1b}
      \ee
      where the phase factor $\theta_{\vec k}$ is at this time arbitrary.
      Note that we have demanded that $|\vec k, \beta>$ be related
      to $|\vec k, \alpha>$ by eq (\ref{PT}).
      In this new basis, we find
      \be
       \mathfrak{m}_z^{\rm eff} (\vec k) = | A_{\vec k}| \rho_z
       \label{mzf}
       \ee
       and
       \begin{widetext}
       \be
       \mathfrak{m}_x^{\rm eff} = \left(
       \ba{cc} 0 & ( |A_{\vec k}| {\rm cos} \phi_{\vec k} + i {\rm sin} \phi_{\vec k})
         e^{ -i (\phi_{\vec k} + \alpha_{\vec k} + 2 \theta_{\vec k})} \\
         ( |A_{\vec k}| {\rm cos} \phi_{\vec k} - i {\rm sin} \phi_{\vec k})
         e^{ i (\phi_{\vec k} + \alpha_{\vec k} + 2 \theta_{\vec k})} & 0 \ea \right)
       \ee

     \be
       \mathfrak{m}_y^{\rm eff} = \left(
       \ba{cc} 0 & ( |A_{\vec k}| {\rm sin} \phi_{\vec k} - i {\rm cos} \phi_{\vec k})
         e^{ -i (\phi_{\vec k} + \alpha_{\vec k} + 2 \theta_{\vec k})} \\
         ( |A_{\vec k}| {\rm sin} \phi_{\vec k} + i {\rm cos} \phi_{\vec k})
         e^{ i (\phi_{\vec k} + \alpha_{\vec k} + 2 \theta_{\vec k})} & 0 \ea \right)
       \ee

       To proceed further, it is simplest to examine the radial and azimuthal components
       of $ \mathfrak{\vec m}^{\rm eff}$ and $\vec \rho$, {\it i.e.}
       $ \mathfrak{m}_r^{\rm eff} \equiv
       {\rm cos} \phi_{\vec k} \mathfrak{m}_x^{\rm eff} +
       {\rm sin} \phi_{\vec k} \mathfrak{m}_y^{\rm eff}$ and
        $ \mathfrak{m}_{\phi}^{\rm eff} \equiv
       - {\rm sin} \phi_{\vec k} \mathfrak{m}_x^{\rm eff} +
       {\rm cos} \phi_{\vec k} \mathfrak{m}_y^{\rm eff}$, and similarly
       for $\rho_r$ and $\rho_{\phi}$, {\it i.e.}
         \end{widetext}
       \bdm
       \rho_r \equiv \left( \ba{cc} 0 & e^{-i \phi_{\vec k}} \\
       e^{i \phi_{\vec k}} & 0 \ea \right)
       \edm
       and
       \bdm
       \rho_{\phi} \equiv \left( \ba{cc} 0 & -i e^{-i \phi_{\vec k}} \\
       i e^{i \phi_{\vec k}} & 0 \ea \right)
       \edm

      Evidently due to the existence of vertical reflection planes at
      arbitrary angles with respect to the x-axis, $\mathfrak{m}_r^{\rm eff}$
      ( $\mathfrak{m}_{\phi}^{\rm eff}$) must simply be proportional
      to $\rho_r$ ( $\rho_{\phi}$) but would not involve the other
      component.  One sees that we can choose $\theta_{\vec k}$ to satisfy
      \be
      \alpha_{\vec k} + 2 \theta_{\vec k} = 0
      \label{theta}
      \ee
      or $\alpha_{\vec k} + 2 \theta_{\vec k} = \pi$.  We shall adopt the first
      choice.  In this case we get
      \be
      \mathfrak{m}_r^{\rm eff} = | A_{\vec k}| \rho_r
      \label{mrf}
      \ee
      and
      \be
      \mathfrak{m}_{\phi}^{\rm eff} = \rho_{\phi}
      \label{mpf}
      \ee
      For this choice, a pseudospin along the positive azimuthal direction would
      correspond also to an effective magnetic moment and hence spin
      along the same direction.
      (The alternate choice would give $\mathfrak{m}_r = - | A_{\vec k}| \rho_r$
      and
      $\mathfrak{m}_{\phi}^{\rm eff} = -\rho_{\phi}$ instead.)
      Back to the Cartesian form, we have
      \be
      \mathfrak{m}_x^{\rm eff} (\vec k) =
      \rho_x - ( 1 - |A_{\vec k}|) \hat k_x (\hat k_x \rho_x + \hat k_y \rho_y)
      \ee
      \be
      \mathfrak{m}_y^{\rm eff} (\vec k) =
      \rho_y - ( 1 - |A_{\vec k}|) \hat k_y (\hat k_x \rho_x + \hat k_y \rho_y)
      \ee
      which explicitly shows that $\mathfrak{m}_{x,y}^{\rm eff}$
      has the same transformation properties as $\rho_{x,y}$ respectively.

      Note that the procedure above also  gives us the effective $g$-factor
      for the effective moments.  For magnetic moment along $z$ and
      the radial component $r$, eq (\ref{mzf}) and (\ref{mrf}) show
      that they are reduced by the factor $|A_{\vec k}| =
      (m^2 + v_z^2 k_z^2)^{1/2}/|E_{\vec k}| < 1 $ given in eq (\ref{Aabs}), but there is no
      reduction for the $\phi$ component.  For $\vec k$ on the Fermi surface,
      $E_{\vec k} = \mu$, $|A_{\vec k}| = 1$ for $\vec k$ parallel or
      antiparallel  to $\hat z$.  It decreases for increasing $k_{\|}$,
      and for $\vec k$ in the x-y plane, $|A_{\vec k}|  \to |m/\mu|$,
      which can be substantially less than unity, as in the case relevant to
      the experiments \cite{Wray10}.   This effective moment would be
      relevant when considering questions such as Pauli limiting
      of upper critical field \cite{Bay12}, or spin susceptibilities
      measured by Knight shifts.
      Returning to the pseudospin basis, eq (\ref{transf1a}) and (\ref{transf1b}) become
       \be
      |\vec k, \alpha> = \frac{e^{-i \alpha_{\vec k}/2}}{\sqrt{2}}
      \left( |\vec k, \alpha'> - i e^{ i (\phi_{\vec k} + \alpha_{\vec k})}
        |\vec k, \beta'> \right)
      \label{kaf}
      \ee
      and
   \be
      |\vec k, \beta> = \frac{e^{i \alpha_{\vec k}/2}}{\sqrt{2}}
      \left( |\vec k, \beta'> - i e^{ -i (\phi_{\vec k} + \alpha_{\vec k})}
        |\vec k, \alpha'> \right)
      \label{kbf}
      \ee
      States at $-\vec k$, $k_z > 0$, can be obtained by using eq (\ref{P}):
     \be
      |-\vec k, \alpha> = \frac{e^{-i \alpha_{\vec k}/2}}{\sqrt{2}}
      \left( |-\vec k, \alpha'> - i e^{ i (\phi_{\vec k} + \alpha_{\vec k})}
        |-\vec k, \beta'> \right)
      \label{-kaf}
      \ee
      and
   \be
      |-\vec k, \beta> = \frac{e^{i \alpha_{\vec k}/2}}{\sqrt{2}}
      \left( |-\vec k, \beta'> - i e^{ -i (\phi_{\vec k} + \alpha_{\vec k})}
        |-\vec k, \alpha'> \right)
      \label{-kbf}
      \ee
      With $|\pm \vec k,\alpha'>$ and $|\pm \vec k, \beta'>$ available in
      eq (\ref{ka1}) (\ref{kb1}), (\ref{-ka1}) (\ref{-kb1}), this
      completes our construction of the pseudospin basis.
      We shall express the Cooper pair wavefunctions in terms of
      it in the next section.

        Before we proceed, since we would also be interested in surface bound
        states in the superconducting states in Sec \ref{sec:surface},
        we consider reflection of quasiparticles at a surface in
        the normal state before we end this section.  We consider a crystal
        occupying $z < 0$, with a surface at $z=0$.
        Consider incident wavevector
        $\vec k = k_x \hat x + k_y \hat y + k_z \hat z$, $k_z > 0$.
        Due to our ways of writing wavefunctions for wavevectors in
        the southern hemisphere,
        it is convenient to write the reflected wavevector as
        $- \vec k'$ where $\vec k' = - k_x \hat x - k_y \hat y + k_z \hat z$
        and use eq (\ref{-kaf}) and (\ref{-kbf}), and note that $\phi_{-\vec k'}
        = \phi_{\vec k}$.  Straight-forward algebra shows that
        $|\vec k, \alpha>$ is reflected only into $|-\vec k', \alpha>$,
        and similarly for $\alpha \to \beta$.  Indeed, the wavefunctions
        $|\Psi> \equiv$ $ |\vec k, \alpha> + R_{\vec k} | - \vec k', \alpha> $ or
        $|\vec k, \beta> + R_{\vec k} | - \vec k', \beta> $,
        with the reflection coefficient
        $R_{\vec k} = - e^{ i \alpha_{\vec k}}$ $= - ({\rm sgn} E_{\vec k})
        \frac{m + i v_z k_z}{ (m^2 + v_z^2 k_z^2)^{1/2}}$ satisfy
        the boundary condition
        $< \sigma_z = -1 | \Psi>  = 0$, since
        one can easily verify that
        $< \sigma_z = -1 | \vec k, \alpha> = e^{ -i \alpha_{\vec k}}
         < \sigma_z = -1 | \vec k', \alpha>$ and similarly with
         $\alpha \to \beta$.
         Hence the reflection of quasiparticles at $z=0$ in the normal state
         does not alter the pseudospin species, nor is the phase shift
         dependent on the incident species.  Hence, in the model of eq (\ref{HN}),
         the $z=0$ surface is {\em pseudospin-inactive}, a result that
         we would use in Sec \ref{sec:surface}.

         Our single band description here is unable to capture the
         surface states of a TI.  These states are superposition of
         states from both the conduction and valence bands.
         The implication of this for the surface states of
         the superconducting phases would be discussed in Sec \ref{sec:surface}.

\section{Superconducting states}

\subsection{Bulk} \label{sec:bulk}

It is straight-forward to obtain the order parameters for the superconducting
states in our pseudospin basis.  We discuss each of the phases listed in
Table \ref{table} in turn.
We confine ourselves to momentum independent pairing within the $\sigma$, $s$
basis of eq (\ref{HN}). Generalization to additional momentum dependence
is straight-forward. So far for Cu:Bi$_2$Se$_3$, superconductivity
has been found only for $\mu> 0$,\cite{Wray10} but we shall also consider general
sign of $\mu$ in the following.

\noindent {$\bf A_{1g}$}:

Both the ``intra-orbital opposite spin pairing" $ | 1 \uparrow, 1 \downarrow>
 + | 2 \uparrow, 2 \downarrow> $ and the ``inter-orbital singlet pairing"
$ | 1 \uparrow, 2 \downarrow> - | 1 \downarrow,  2 \uparrow> $ has $A_{1g}$ symmetry.
In general, they are expected to be mixed.  However, since they
have also been discussed separately in, e.g., \cite{Hao11}, we shall
also first do the same likewise, and consider a general linear combination later.
For ease of referral, we shall refer these two states as
$A'_{1g}$ and $A^{''}_{1g}$ respectively.

\noindent $A'_{1g}$:

  The pair wavefunction is  $ | 1 \uparrow, 1 \downarrow>
 + | 2 \uparrow, 2 \downarrow> $.  The corresponding form in the pseudospin
 language is just
 $\sum'_{\vec k; \gamma, \gamma' = \alpha, \beta}
 | \vec k \gamma, -\vec k \gamma' >
 \left[ < \vec k \gamma, -\vec k \gamma' | 1 \uparrow, 1 \downarrow >
   + < \vec k \gamma, -\vec k \gamma'| 2 \uparrow, 2 \downarrow> \right] $
   Here the prime over the sum means that $\vec k$ is restricted
   to the "upper hemisphere" (as those in the other hemisphere
   is already included by antisymmetry),  and
   $ < 1 \uparrow, 1 \downarrow | \vec k \gamma, -\vec k \gamma' >
   = < 1 \uparrow |\vec k \gamma>< 1 \downarrow | -\vec k \gamma'>
     - < 1 \downarrow | \vec k \gamma> < 1 \uparrow|  -\vec k \gamma'>$
     can be evaluated using (\ref{kaf})-(\ref{-kbf}) and
      eq (\ref{ka1})-(\ref{-kb1}).
      We find that this simply reduces to
      $\sum'_{\vec k} \left[ | \vec k \alpha, - \vec k \beta> -
       | \vec k \beta, - \vec k \alpha> \right] $, with no
       additional momentum dependent factors.  If the pairing term
       in the superconducting Hamiltonian is taken as
       $ \Delta'_1 ( c^{\dagger}_{1 \uparrow} c^{\dagger}_{1 \downarrow}
          + c^{\dagger}_{ 2 \uparrow} c^{\dagger}_{2 \downarrow}) + h.c.$
          where $c^{\dagger}_{\sigma,s}$ are the creation operators
          of orbit $\sigma$ and spin $s$ and $h.c.$ indicates the
          Hermitian conjugate, then the corresponding term in the pseudopin
          basis simply reads $\sum'_{\vec k} \Delta'_1 ( c^{\dagger}_{\vec k \alpha}
          c^{\dagger}_{-\vec k \beta} - c^{\dagger}_{\vec k \beta} c^{\dagger}_{-\vec k \alpha})
          + h.c.$.  This has just the familiar form for momentum independent
          conventional $s$-wave pairing.  The quasiparticle energies in the
          superconducting state, measured with respect to the chemical potential
          $\mu$, are $\pm E_S$ with $E_S$ in the familiar form,
          for $\mu > 0$
          \be
          E^2_S = (\epsilon_{\vec k} - \mu )^2 + (\Delta'_1)^2
          \label{ESA11}
          \ee
          where we have taken the gauge where $\Delta_1$ is real.
          The corresponding formula for $\mu < 0$ is
          \be
          E^2_S = (\epsilon_{\vec k} + \mu )^2 + (\Delta'_1)^2 \ .
          \label{ESA11n}
          \ee
          Note that since we started with a normal metal and then
          introduce the superconducting pairing,  we necessarily
          have $\mu^2 > m^2$ implicitly, and weak-superconducting
          pairing actually requires further that
          $|\mu| - |m| \gg |\Delta'_1|$.  For the full two-band description,
          the Hamiltonian in the Nambu-II notation is just
          $(H_N - \mu) \tau_z + \Delta'_1 \tau_x$ (see Table \ref{table}),
          where $\tau_{x,y,z}$ are the Pauli matrices in particle-hole space.
          Now there are instead two pairs of allowed $E_S$ due to
          the presence of two bands, given by
          $ E^2_S = (\epsilon_{\vec k} \pm \mu )^2 + (\Delta'_1)^2$,
          thus including both eq (\ref{ESA11}) and (\ref{ESA11n})
          irrespective of the sign of $\mu$.

 \noindent $A^{\prime\prime}_{1g}$:

   The pair wavefunction is $ | 1 \uparrow, 2 \downarrow > - | 1 \downarrow, 2 \uparrow>$.
   Following the same procedure described above  gives
   the result $\sum'_{\vec k} \frac{m}{E_{\vec k}}
      (| \vec k \alpha, - \vec k \beta> - |\vec k \beta, -\vec k \alpha >) $
      in the pseudospin basis.
      If the pairing term in the Hamiltonian is written
      as
      $\Delta^{''}_{1} ( c^{\dagger}_{1 \uparrow} c^{\dagger}_{2 \downarrow}
       - c^{\dagger}_{1 \downarrow} c^{\dagger}_{2 \uparrow}) + h.c.$
       with $\Delta^{''}_1$ real,
        the corresponding quasiparticle energies are $\pm E_S$
        with
        \be
        E_S^2 = (\epsilon_{\vec k} \mp \mu)^2 + \left(\frac{m}{\mu}\Delta^{''}_1 \right)^2
        \label{EA11}
        \ee
        with the upper and lower signs for $\mu> 0$ and $\mu < 0$ respectively.
        Again the energy gap is isotropic in $\vec k$ space and is
        given here simply by $| \frac{m}{\mu}\Delta^{''}_1 |$.
        Note that the system becomes gapless if $m=0$ even for
        finite $\Delta^{''}_1$.
        On the other hand, in the full two-band description,
        with Hamiltonian $H_S = ( H_N - \mu) \tau_z + \Delta^{''}_1 \sigma_x \tau_x$,
        we obtain
        again two pairs of energies, given by
        \be
        E_S^2 = \epsilon_{\vec k}^2 + \mu^2 + (\Delta^{''}_1)^2
         \pm 2 \left[ \mu^2 \epsilon_{\vec k}^2 + (\Delta^{''}_1)^2 (\mu^2 - m^2)
         \right]^{1/2}
         \ee
         Considering the lower energy branch (where $\epsilon_{\vec k} \approx
         \pm \mu$ for $\mu {> \atop <} 0$) and
         taking the weak-pairing $|\mu| - |m| \gg |\Delta_1|$ approximation,
         we recover eq (\ref{EA11}), as expected.
         Generally, the system is gapped whenever $m \ne 0$ and $\Delta^{''}_1 \ne 0$.
         If $\Delta^{''}_1 = 0$, we recover the normal state, and
         states at momentum $\vec k$ such that $\epsilon_{\vec k} = \pm \mu$
         have $E_S = 0$.
         If $m= 0$, gaplessness occurs if  $\epsilon_{\vec k}^2 = \mu^2 + (\Delta^{''}_1)^2$
         can be satisfied, and at
         positions slightly different from the one-band result
         $\epsilon_{\vec k} = \pm \mu$ with correction due to finite $\Delta^{''}_1$.

 \noindent general $A_{1g}$:

 The most general $A_{1g}$ pairing wavefunction is a linear combination
 of that of $A'_{1g}$ and $A^{''}_{1g}$.  The general
 pairing Hamiltonian is
     $\Delta^{'}_{1} ( c^{\dagger}_{1 \uparrow} c^{\dagger}_{1 \downarrow}
       + c^{\dagger}_{2 \uparrow} c^{\dagger}_{2 \downarrow}) +
           \Delta^{''}_{1} ( c^{\dagger}_{1 \uparrow} c^{\dagger}_{2 \downarrow}
       - c^{\dagger}_{1 \downarrow} c^{\dagger}_{2 \uparrow}) + h.c.$.
       The corresponding expression in pseudospin language is again
       just the linear combination of those given above for
       $A'_{1g}$ and $A^{''}_{1g}$.
       We shall, for simplicity, restrict ourselves only to the
       case where time-reversal symmetry is preserved, thus
       $\Delta'_1$ and $\Delta^{''}_1$ can be chosen to be real simultaneously,
       though either can be positive or negative.
       The energy spectrum is, in the single band description
       \be
       E_S^2 = ( \epsilon_{\vec k} \mp \mu)^2 +
       \left( \Delta'_1 + \frac{m}{\mu} \Delta^{''}_1 \right)^2
       \label{ES1g}
       \ee
       The system is gapped if $ \Delta'_1 + \frac{m}{\mu} \Delta^{''}_1 \ne 0$.
       The corresponding result in the full two-band description is
       \begin{widetext}
       \be
       E_S^2 = \epsilon_{\vec k}^2 + \mu^2 + (\Delta'_1)^2 + (\Delta^{''}_1)^2
       \pm 2 \left[ \mu^2 m^2 + (\Delta'_1 \Delta^{''}_1)^2
        - 2 \mu m \Delta'_1 \Delta^{''}_1 + (\epsilon_{\vec k}^2 - m^2 )
        ( \mu^2 + \Delta^{'' 2}_1 ) \right]^{1/2}
        \label{ES1f}
        \ee
       \end{widetext}
      The lower energy branch reduces to eq (\ref{ES1g}) in the weak-pairing limit.
      If the full expression (\ref{ES1f}) is used, one can check that
      the system is gapped whenever $m \Delta^{''}_1 + \Delta^{'}_1 \mu \ne 0$.
      (Thus recovering the one band result since there $\mu$ must be finite).
      If $m \Delta^{''}_1 + \Delta^{'}_1 \mu = 0$, gaplessness still requires
      the condition $ (\vec v \cdot \vec k)^2 = (\mu^2 - m^2)
        - (\Delta^{'2}_1 - \Delta^{'' 2}_1)$ be satisfied.
        Hence we have the following :

        (1):  If $\mu \ne 0$, we need $ \Delta'_1 + \frac{m}{\mu} \Delta^{''}_1 = 0$ and
        $ (\vec v \cdot \vec k)^2 = (\mu^2 - m^2) \left( 1 -
         (\Delta^{''}_1/{\mu})^2 \right)$
        for gaplessness.  Note that for $\mu^2 {> \atop <} m^2$,
        the latter happens if and  only if $\mu^2 {> \atop < } \Delta^{'' 2}_1$.
        We reproduce the weak superconducting pairing results when
        $|\mu| - |m| \gg \Delta'_1$ and $\Delta^{''}_1$.

        (2) For $\mu = 0$, the system is gapped whenever $m \ne 0$ and $\Delta^{''}_1 \ne 0$.
        (2a) If $m = 0$, gaplessness occurs only if $|\Delta^{''}_1| > |\Delta'_1|$.
        (2b) If $\Delta^{''}_1 = 0$, the system reduces to $A'_{1g}$, and the
        system is fully gapped unless $\Delta'_1$ also vanishes.
        We shall use these results when we discuss the surface states and topology in
        Sec \ref{sec:surface} and the Appendix.

        \noindent ${\bf A_{1u}}$:  The pairing wavefunction,
        $ | 1 \uparrow, 2 \downarrow > + | 1 \downarrow, 2 \uparrow >$
        becomes
        \begin{widetext}
        \bdm
        \frac{m}{ ( m^2 + v_z^2 k_z^2 )^{1/2} }
        \sum'_{\vec k} \left[
         - \frac {v (k_x - i k_y )} {|E_{\vec k}|}| \vec k \alpha, - \vec k \alpha >
          + \frac {v (k_x + i k_y )} {|E_{\vec k}|} | \vec k \beta, - \vec k \beta >
          + ({\rm sgn} E_{\vec k}) \frac{ v_z k_z }{m}
          (| \vec k \alpha, - \vec k \beta > + | \vec k \beta, - \vec k \alpha >) \right]
          \edm
          \end{widetext}
          If the pairing term in the Hamiltonian is written as
          $\Delta_{1u} (c^{\dagger}_{1 \uparrow} c^{\dagger}_{2 \downarrow}
       + c^{\dagger}_{1 \downarrow} c^{\dagger}_{2 \uparrow}) + h.c.$,
       with $\Delta_{1u}$ real and positive,
       the corresponding form in pseudospin is
       $ ( - d_x + i d_y ) c^{\dagger}_{\vec k \alpha} c^{\dagger}_{-\vec k \alpha}
         + d_z (c^{\dagger}_{\vec k \alpha} c^{\dagger}_{-\vec k \beta}
           + c^{\dagger}_{\vec k \beta} c^{\dagger}_{-\vec k \alpha})
            + ( d_x + i d_y ) c^{\dagger}_{\vec k \beta} c^{\dagger}_{-\vec k \beta}$
            with, when quasiparticles are taken at the Fermi energy $\mu$,
            \be
            d_{x,y} = \Delta_{1u} \frac{m}{|\mu|} \frac{v k_{x,y}}{ ( m^2 + v_z^2 k_z^2 )^{1/2} }
            \label{dA1u}
            \ee
            \be
            d_z = \Delta_{1u} ({\rm sgn} \mu) \frac{v_z k_z}{ ( m^2 + v_z^2 k_z^2 )^{1/2} }
            \label{dzA1u}
            \ee
            As compared with the Balian and Werthamer (BW) state \cite{BW}
            where $\vec d (\vec k) \| \vec k$ (or $\vec d (\vec k)$
            anti-parrallel to $\vec k$, after a gauge transformation),
             the $\vec d(\vec k)$ here
            has a very peculiar form.  The ratio between the in-plane and
            $z$ component is
            \be
            \frac{d_\|}{d_z} = \frac{m}{\mu} \frac{v k_{\|}}{v_z k_z}
            \label{dratio}
            \ee
            Besides the anisotropy factors from the velocities, an extra
            factor $m/\mu$ arises, which suppresses the $d_\|$ component
            relative to $d_z$ if $|m| < |\mu|$.  This, in retrospect,
            is  actually not surprising since the pairing is between
            opposite spins in the original $\sigma$ and $s$ basis.
            A pure opposite pseudospin pairing would have $\vec d$
            parallel or antiparallel to $\hat z$.
            The $x,y$ components of $\vec d$ are actually generated by
            spin-orbit coupling.
            Moreover, we note that the relative signs between
            $d_{x,y}$ and $d_z$ depends on the signs of
            the various parameters of the system, in particular
            ${\rm sgn} (v_z m)$.  We shall come back to this
            when we discuss the surface bound states.
            We note here also that this peculiar relative sign
            and magnitudes between the components of $\vec d$
            is allowed here due to the inequivalence between
            $z$ and $x$-$y$ under the relevant $D_{3d}$ symmetry.
            For a cubic system such as YPtBi, \cite{Butch11}
            $k_x \hat x + k_y \hat y$ and $k_z \hat z$ necessarily
            comes in the combination
            $k_x \hat x + k_y \hat y + k_z \hat z$.

            Despite the peculiar form for $\vec d$, the energy
            gap turns out to be isotropic in the weak-coupling limit.
            The square of this gap is given by $\vec d \cdot \vec d$,
            which is
            \bdm
            \frac{\Delta_{1u}^2}{ m^2 + v_z^2 k_z^2 }
            \left[ \left(\frac{m v k_\|}{\mu}\right)^2 + v_z^2 k_z^2 \right]
            \edm
            which works out to be simply $\Delta_{1u}^2 ( 1 - (m/\mu)^2 )$
            when we restrict ourselves to
            particles near the Fermi surface, where
            $m^2 + v_z^2 k_z^2 + v_\|^2 k_\|^2  = \mu^2$.
            For the energy gap, the anisotropies due to eq (\ref{dratio})
            and the overall factors $(m^2 + v_z^2 k_z^2)^{1/2}$
            in eq (\ref{dA1u}) and (\ref{dzA1u}) cancel each other.
            The quasiparticle energies are just
            $(\epsilon_{\vec k} \mp \mu)^2 + \Delta_{1u}^2 ( 1 - (m/\mu)^2 )$.
            This phase is fully gapped (provided $\mu^2 > m^2$, which,
            as mentioned,  is necessarily the case for a single-band weak-pairing
            superconductivity picture to be meaningful).

            In the full two-band description, the quasiparticle energies
            are already worked out in \cite{FuBerg10}:
            \be
            E_S^2 = \epsilon_{\vec k}^2 + \mu^2 + \Delta_{1u}^2
             \pm 2 \left[ \mu^2 \epsilon_{\vec k}^2 + m^2 \Delta_{1u}^2 \right]^{1/2}
             \label{ESA1u}
             \ee
             which reduces to what has been just given in the weak-pairing limit.
             In this two-band description, the system can still be
             gapless when $\Delta_{1u} \ne 0$ provided
             $m = \pm \sqrt{ \mu^2 + \Delta_{1u}^2}$ (with gapless point
             at $k=0$).  We shall use this
             result later in the Appendix.

             \noindent $\bf A_{2u}$:

             The pair wavefunction $ | 1 \uparrow, 1 \downarrow>
              - | 2 \uparrow, 2 \downarrow >$ becomes
              $i \sum'_{\vec k} \left[ \frac{v (k_x - i k_y)}{E_{\vec k}} |\vec k \alpha, - \vec k \alpha>
                + \frac{v (k_x + i k_y)}{E_{\vec k}} |\vec k \beta, - \vec k \beta> \right] $.
             If the pairing term is written as  $\Delta_{2u}
             (c^{\dagger}_{1 \uparrow} c^{\dagger}_{1 \downarrow}
            -   c^{\dagger}_{2 \uparrow} c^{\dagger}_{2 \downarrow}) + h.c.$,
            the corresponding
             $\vec d (\vec k)$ is, in the weak pairing limit,
             \be
             \vec d (\vec k) = \Delta_{2u} \frac{v(k_x \hat y - k_y \hat x)}{\mu}
             \label{dA2u}
             \ee
            The magnitude of the gap is just $\Delta_{2u} | v k_\| / \mu |$.
            This is the usual planar phase in the $^3$He literature,
            and is regaining attention due to its analogy with topological
            insulators in two-dimensions (e.g. \cite{Schnyder08,Yip10}).
            In  three-dimension however, this state has point nodes in the
            gap at the north and south poles of the Fermi surface,
            where $k_x$, $k_y$ both vanish.

            The expression for the quasiparticle energies in the two-band
            description is
            \be
            E_S^2 = \epsilon_{\vec k}^2 + \mu^2 + \Delta_{2u}^2
            \pm 2 \left[ \mu^2 \epsilon_{\vec k}^2 + \Delta_{2u}^2 ( m^2 + v_z^2 k_z^2) \right]^{1/2}
            \label{ESA2u}
            \ee
            which reduces to the above results in the weak-pairing limit.
            The state is gapped at all $v k_\| \ne 0$.  Gaplessness can
            occur if the condition $v_z^2 k_z^2 = \mu^2 + \Delta_{2u}^2 - m^2 $
            can be satisfied.

            \noindent $\bf E_{u}$:

            This is a two-dimensional representation, as $|1 \downarrow 2 \downarrow>$
            is the time-reversed of $ | 1 \uparrow 2 \uparrow >$.  Generally,
            the superconducting state can be a superposition of the two.
            Let us first consider the state $ i | 1 \uparrow, 2 \uparrow > $
            (we have inserted an $i$ factor for later convenience.)
            If the pairing term in the Hamiltonian is given by
            $ i \Delta_{u} ( c^{\dagger}_{1 \uparrow} c^{\dagger}_{2 \uparrow}) + h.c.$,
            then we have
            \begin{widetext}
            \bea
            \vec d (\vec k) &=& \Delta_u
            \left\{
             \frac{1}{4} ({\rm sgn} \mu)
            \left( \frac{1}{ (m^2 + v_z^2 k_z^2)^{1/2}} + \frac{1}{|\mu|} \right)
            v_z k_z \hat r_+
             -  \frac{m}{ (m^2 + v_z^2 k_z^2)^{1/2}} \frac{v k_+}{ 2 |\mu|}  \hat z
                    \nonumber \right. \\
              & & \left. +  \frac{1}{4} ({\rm sgn} \mu)
            \left( \frac{1}{ (m^2 + v_z^2 k_z^2)^{1/2}} - \frac{1}{|\mu|} \right) v_z k_z
               \left( \cos (2 \phi_{\vec k}) \hat x + i \sin (2 \phi_{\vec k}) \right) \hat r_-
               \right\}
               \label{dEu}
               \eea
               \end{widetext}
            This $\vec d (\vec k)$ is complex ($\vec d \times \vec d^* \ne 0$) reflecting
            the fact that the state $ | 1 \uparrow 2 \uparrow >$ has broken time-reversal symmetry.
            The first two terms in eq (\ref{dEu}) are proportional to
            $k_z \hat r_+$ and $k_+ \hat z$ listed under $E_u$ in Table \ref{table}.
            The last term has a more complicated momentum dependence, but
            since
            \bdm
            \left[ \frac{1}{ (m^2 + v_z^2 k_z^2)^{1/2}} - \frac{1}{|\mu|} \right]
             \approx \frac{1}{2} \frac{ v^2 k_\|^2}{\mu^2}
             \edm
             for small $k_\|$, it is simply proportional to $k_z k_+^2 \hat r_-$,
             the third independent basis function listed in Table \ref{table}, in this limit.
             The spectrum for this state is complicated since it is "non-unitary",
             that is, the energy of the two pseudospin-species at the same $\vec k$ point
             are typically unequal, due to the lack of time-reversal symmetry.
             We shall not investigate this phase in detail, but turn to
             the time-reversal symmetric states within this two-dimensional manifold.

             Let us consider then $i | 1 \uparrow, 2 \uparrow > - i | 1 \downarrow, 2 \downarrow>$.
             This state is just the linear combination of the one discussed above and its
             time-reversal conjugate.  The $\vec d (\vec k)$ vector for this
             state is therefore simply twice the real part of eq (\ref{dEu}),
             and so
             \begin{widetext}
            \bea
            \vec d (\vec k) = \Delta_u
            \left\{
            \frac{1}{2} ({\rm sgn} \mu)
            \left( \frac{1}{ (m^2 + v_z^2 k_z^2)^{1/2}} + \frac{1}{|\mu|} \right)
            v_z k_z \hat x
             -  \frac{m}{ (m^2 + v_z^2 k_z^2)^{1/2}} \frac{v} {|\mu|} k_x \hat z \right.
                    \nonumber \\
                 + \left. \frac{1}{2} ({\rm sgn} \mu)
            \left( \frac{1}{ (m^2 + v_z^2 k_z^2)^{1/2}} - \frac{1}{|\mu|} \right) v_z k_z
               \left( \cos (2 \phi_{\vec k}) \hat x +  \sin (2 \phi_{\vec k}) \hat y \right)
              \right\}
               \label{dEu1}
               \eea
               \end{widetext}
               corresponding to the basis functions listed in the first line under
               $E_u$ in Table \ref{table}.
               Despite its complicated form,  the square of the gap, obtained from
               $\vec d \cdot \vec d$, is given simply by
               \bdm
               \Delta_u^2 \frac{v^2 k_x^2 + v_z^2 k_z^2}{\mu^2}
               \edm
               This state has two point nodes, as it is
               gapless for $\vec k$ parallel to $\hat y$.  The result is in
               accordance with the full two-band result,  which is
               \be
               E_S^2 = \epsilon_{\vec k}^2 + \mu^2 + \Delta_u^2
                \pm 2 \left[ \mu^2 \epsilon_{\vec k}^2 + \Delta_u^2 (m^2 + v^2 k_y^2) \right]^{1/2}
                \ee
                and is just eq (\ref{ESA2u}) with $v_z k_z \to v k_y$.
                The gap-squared in the weak-coupling limit is
                \bdm
                \Delta_u^2 \left[ 1 - \frac{m^2 + v^2 k_y^2}{\mu^2} \right]
                 = \Delta_u^2 \frac{v^2 k_x^2 + v_z^2 k_z^2}{\mu^2}
                \edm
                for momenta on the Fermi surface.  The point node for this
                phase has also been noted in \cite{Yamakage12}.

                $\vec d (\vec k)$  for the state $| 1 \uparrow 2 \uparrow>
                + | 1 \downarrow 2 \downarrow>$ is evidently twice
                the imaginary part of eq (\ref{dEu}).  It is just
                eq (\ref{dEu1}) rotated by $\pi/2$ about $\hat z$.

\subsection{surface states} \label{sec:surface}

              Now we consider the surface states in the superconducting phases.
              We shall focus on the odd parity states since they have
              received more attention in the literature. (see however the
              Appendix)

              For weak superconductors (pairing potential much smaller than
              fermi energy), surface bound states are most conveniently discussed
              quasiclassically.  Bound states can be formed at the surface
              since the quasiparticle with incident wavevector $\vec k_{\rm in}$
              sees a different order parameter from when it is reflected into
              $\vec k_{\rm out}$.  Since the surface $z=0$ is  pseudospin inactive
              \cite{phaseshift}, the problem maps to the evaluation of
              the quasiparticle bound states at a one-dimensional junction,
              where the order parameter for $z'< 0$ is different from that
              for $z'> 0$. Here, the order parameter for $z'< 0$ can be identified
              with that of $\vec k_{\rm in}$, and $z'> 0$ with $\vec k_{\rm out}$.
              In all situations relevant to us, the magnitude of the order parameter
              $|\Delta'|$ for
              $\vec k_{\rm in}$ and $\vec k_{\rm out}$ are identical, only
              the phase $\zeta$'s are different.  The effective phase difference
              for the junction is $\chi = \zeta_{\rm out}- \zeta_{\rm in}$.
              The absolute value of the bound state energy is just
              $|E_b| = |\Delta'| |\cos (\chi/2)|$ if assume that
              the order parameter is constant up to the surface.
              Since we shall be interested also in the sign of $E_b$,
              we give a short discussion of it \cite{Eb}.  For normal state
              with particle-like dispersion, where the energy is increasing with
              the magnitude of the wavevector, the bound state energy
              for positive $z'$ momentum is $E_b = - |\Delta'| \cos (\chi/2) < 0$
              for $0 < \chi < 2 \pi$, and $E_b = |\Delta'| \cos (\chi/2) > 0$
              for $ - 2 \pi < \chi < 0$, and vice versa for negative $z'$ momentum.
              The above signs should be reversed for hole-like normal
              state dispersions \cite{sign}.
              For our normal state, the spectrum is particle-like
              for $\mu > 0$, and hole-like if $\mu < 0$.

              For time-reversal symmetric odd parity superconductors,
              $\vec d (\vec k)$ can be chosen real.  It is best to work
              with the quantization axes $\hat z'_s$ (no relation
              to $z'$ above) which is perpendicular to both
              $\vec d_{\rm in} \equiv \vec d (\vec k_{\rm in})$ and
              $\vec d_{\rm out} \equiv \vec d (\vec k_{\rm out})$.
              We shall choose $\hat z'_s$ to be parallel to
              $\vec d_{\rm in} \times \vec d_{\rm out}$.
              Then the order parameter with $\hat z'_s$ as the
              quantization axis is given by
              \bdm
              \left( \ba{cc} - d_{x'_s} + i d_{y'_s} & 0 \\
              0  & d_{x'_s} + i d_{y'_s}
              \ea \right) \edm
              hence diagonal in pseudospin space.
              Consider first the "down" component.  We can write
              $(d_{x'_s} + i d_{y'_s})_{\rm in/out} = |\Delta'| e^{\zeta_{\rm in/out}}$
              where $\zeta_{\rm in/out}$ is just the angle for $\vec d_{\rm in/out}$
              in the $x_s$ - $y_s$ plane, measured counterclockwise
              from the $x_s$ axis.  Hence the phase difference
              $\chi$ for the ``junction" is just $\chi = \zeta_{\rm out} - \zeta_{\rm in}
               \equiv \zeta$, the angle between $\vec d_{\rm in}$ and $\vec d_{\rm out}$,
               with $ 0 < \zeta < \pi$, and with $E_b = - |\Delta'| \cos (\zeta/2) < 0$
               for particle-like normal state spectrum.
               For pseudospin "up" along $\hat z'_s$, we can rewrite
               $ (- d_{x'_s} + i d_{y'_s})_{\rm in/out} = - |\Delta'| e^{-\zeta_{\rm in/out}}$,
               and the effective phase difference is
               $\chi = ( - \zeta_{\rm out}) - ( - \zeta_{\rm in}) = - \zeta$.
               The bound state energy is
                $E_b =  |\Delta'| \cos (\zeta/2) > 0$ for a particle-like normal state spectrum.

               Summarizing, the bound state energy is positive
               (negative) if the pseudospin is parallel (antiparallel) to
               $({\rm sgn} \mu) \vec d_{\rm in} \times \vec d_{\rm out}$.
               We are particularly interested in comparing this sign with
               the bound state in the normal phase of our TI.  We recall that,
               in our model, the energy is positive if the spin is parallel
               $v \hat n \times \vec k$, where $\hat n$ is the surface normal
               (pointing outward from sample).
               Hence, if we focus on the
               relative sign between the superconducting and the normal
               state, we can state that:

               \noindent The bound state dispersion for the normal
               and superconducting phases has the same sign if
               $({\rm sgn} \mu) \vec d_{\rm in} \times \vec d_{\rm out}$ is
               parallel to $v \hat n \times \vec k$.
               \\

               We shall call this situation as "regular relative to normal" (RN).
               Conversely, we shall call it "anomalous relative to normal" (AN).

               Actually, another meaningful comparison would be to the surface
               bound state for the BW phase where $\vec d$ is parallel to $\vec k$.
               In that case, then the dispersion has the same (opposite) sign as the
               BW phase if $\vec d_{\rm in} \times \vec d_{\rm out}$ is
               parallel (antiparallel) to $\vec k_{\rm in} \times \vec k_{\rm out}$.
               Since $\vec k_{\rm in}$ and $\vec k_{\rm out}$ differ only
               by the component along the surface normal, we see that
               $\vec k_{\rm in} \times \vec k_{\rm out}$ is simply parallel to
               $\hat n \times \vec k_{\rm in}$.  We shall mainly be
               focusing on the first comparison, though the comparison
               with the BW phase can be directly read-off from the expressions
               below.

               Before we proceed further, we remark here that the above single band
               argument only takes into account bound states formed by superposition
               of particle and holes of the same band, with energies
               close to the Fermi level.  For the TI in its normal
               phase however, there are bound states at $E = \pm v k_\|$
               formed by superposition of the conduction and valence band.
               Hence, our single band approximation for superconducting bound states
               is applicable only
               when the states at $E = \pm v k{\|}$ are sufficiently far
               away from $\mu$ so that we can ignore the hydridization
               of our states with these due to the superconducting pairing $\Delta$,
               {\it i.e.}, we need $|\mu| - | v k_\| | \gg \Delta$.
               For $|\mu| \gg m$, we thus
               we expect that we can capture the superconducting bound
               states only when $k {< \atop \sim} k_{F \|}$,
               the Fermi momentum.  More discussion on this will be given below.

               Now we apply our above results to the odd-parity phases.

               \noindent $A_{1u}$: $\vec d (\vec k)$ is available in
               eq (\ref{dA1u}) and (\ref{dzA1u}). We write $ \vec k_{\rm in}
                = k_x  \hat x + k_y \hat y + k_z \hat z$ and
                  $ \vec k_{\rm out}
                = k_x  \hat x + k_y \hat y - k_z \hat z$, $k_z > 0$.
                Since the system is rotationally symmetric about $\hat z$,
                let us consider $k_y = 0$.  We get
                \be
                ({\rm sgn} \mu) \vec d_{\rm in} \times \vec d_{\rm out}
                 = 2 | \Delta_{1u}|^2
                 \frac{ m v_z k_z v k_x}{ (m^2 + v_z^2 k_z^2) |\mu|} \hat y
                 \ee
                 with $v \hat n \times \vec k =  v k_x \hat y$.
                 We see that the dispersion is RN if ${\rm sgn} (m v_z) > 0$,
                 but AN if ${\rm sgn} (m v_z) < 0$.
                 The magnitude of the group velocity of the bound state,
                 $ | d E_b / d k_{\|}  |$, can also be obtained easily.
                 For small $k_x$, ${\rm sin} \zeta =
                 \vec d_{\rm in} \times \vec d_{\rm out} / | \Delta'|^2
                 \approx ( \pi - \zeta) $, since $\zeta$ is close to $\pi$,
                 so $|E_b| \approx |\Delta'| ( \pi - \zeta) /2$.
                 Using $|\Delta'| = \Delta_{1u} ( 1 - (m/\mu)^2 )^{1/2}$,
                 we get
                 \be
                  | d E_b / d k_x | = |\Delta_{1u}| \frac{|m|}{|\mu|}
                   \left( 1 - \left(\frac{m}{\mu}\right)^2 \right)^{-1/2}
                   \frac{v k_x} {|\mu|}
                   \ee
                   For $|m/\mu| \ll 1$, this group velocity is reduced
                   compared with the BW phase by a factor $|m/\mu|$
                   (see also \cite{HsiehFu12}).

                   In our single band description, the bound state energy
                   approaches the bulk gap when $k_\|$ approaches
                   the Fermi momentum $k_{F \|}$, here given by
                   $ (\mu^2 - m^2 )^{1/2}/|v|$, as the effective phase
                   difference $\chi$ vanishes when
                   $\vec k_{\rm in}$ and $\vec k_{\rm out}$ becomes parallel.
                   The bound state spectrum is thus of the form in Fig \ref{fig1}(a)
                   or Fig \ref{fig1}(b) according to whether ${\rm sgn} (m v_z)
                   {> \atop < } 0$ (when $v$ taken as $< 0$ according to
                   Sec \ref{sec:back}).  In contrast, the bound state spectrum
                   in the full two-band calculation for ${\rm sgn} (m v_z) < 0$
                   is schematically shown in Fig \ref{fig1} (c) \cite{Hao11,HsiehFu12}.
                   Though our one-band model captures correctly the sign
                   of the group velocity for  $k_\| {< \atop \sim} k_{F \|}$,
                    it does not capture the behavior
                   at larger $k_\|$.  This is in retrospect not surprising.
                   For large $k_\|$, the Cooper pairing plays no role,
                   and the sign for the dispersion must be the same
                   as the corresponding normal phase.  In the later case,
                   when the system is a TI, ${\rm sgn} (m v_z) < 0$ and
                   the positive energy branch has spin along $v \hat z \times \vec k$.
                   Hence, for ${\rm sgn} (m v_z) < 0$, the sign of
                   the dispersion at small and large $k_{\|}$ must be opposite,
                   as in Fig \ref{fig1}(c).  This sign change
                   has been found earlier by other authors \cite{Hao11,HsiehFu12,Yamakage12}.
                    An alternative view of this sign change has been given by \cite{HsiehFu12}.
                   (see also Appendix \ref{sec:app} for more discussions.)

  \noindent $A_{2u}$:  $\vec d(\vec k)$ is given in eqn (\ref{dA2u}),
  which is independent of the sign of $m$ and $v_z$. Since $\vec d (\vec k)$
  is independent of $k_z$, $\vec d_{\rm in} = \vec d_{\rm out}$.
  There are no surface bound states  (in our approximation where
  the order parameter is constant up to the surface).  This conclusion
  is in agreement with Fig 5 a, c of \cite{Hao11}.  Note that
  the gap magnitude vanishes for normal incidence, where
  $k_x = k_y = 0$.

  \noindent $E_{u}$.  We consider the state $ i ( | 1 \uparrow, 2 \uparrow>
  - | 1 \downarrow, 2 \downarrow >) $, the first line under $E_{u}$
  in Table \ref{table}.  (The result for the second line is the same
  except for a $\pi/2$ rotation about $z$).
  $\vec d (\vec k)$ is available in eq (\ref{dEu1}).  For near normal
  incidence, it can be rewritten as

            \begin{widetext}
            \bea
            \vec d (\vec k) = \Delta_u
            \left\{
              - \frac{m (v k_x)}{|\mu|^2} \hat z +
             ({\rm sgn} \mu) \frac{v_z k_z}{|\mu|}
             \left[ 1 + \frac{1}{4} \frac{ v^2 (k_x^2 - k_y^2)}{\mu^2} \right] \hat x
             + ({\rm sgn} \mu) \frac{v_z k_z}{2 |\mu|}
              \frac{ v^2 k_x k_y}{\mu^2} \hat y \right\}
               \eea
               \end{widetext}
  The factor $v^2 (k_x^2 - k_y^2)/\mu^2$ only gives a small correction
  to the results below and will be ignored for simplicity.  For the $k_x$-$k_z$ plane,
  $k_y =  0$,
  \be
  ({\rm sgn} \mu) \vec d_{\rm in} \times \vec d_{\rm out}
   = 2 \Delta_u^2  \frac{ m v_z k_z  v k_x }{ |\mu|^3} \hat y
   \ee
   whereas $v \hat n \times \vec k = v k_x \hat y$.
   Thus we have RN if ${\rm sgn} (m v_z) > 0$ and
   AN if vice versa.  For the later case,  as argued in
   the last paragraph for $A_{1u}$, the dispersion should be the
   same as the normal phase for large $k_{\|}$, hence
   we again expect a sign change for the group velocity at
   some $k_{\|}$.  This is in accordance with Fig 6a of \cite{Hao11}
   (recall \cite{compare})
   and of \cite{Yamakage12}.

   Before we depart from this section, we would like to make a remark
   on the more general case where the momentum dependence
   of the term $m$ in eq (\ref{HN}) is included.  All the above
   single-band calculations can be simply generalized to this case.
   For  weak superconducting pairing, it is the value of $m(\vec k)$
   at the Fermi surface that is physically relevant.  For a TI,
   $m(\vec k) = m_0 + C k^2$ where $m_0$ and $C$ has opposite
   signs (We shall ignore possible anisotropies in the $C$ term
   as they do not affect the arguments below).  Thus it is
   possible that the sign of $m(\vec k)$ for $\vec k$ at the Fermi surface
   be different from $m(\vec k = 0) = m_0$.  Hence,
   ${\rm sgn} (m (\vec k) v_z)$ can be positive for a TI,
   even though ${\rm sgn} (m_0 v_z) < 0$ is required,
   and the spectra can change from AN to RN with increasing
   $|\mu|$ or $|C|$ (for $A_{1u}$ and $E_{u}$),
   at the point where $m_0 + C \frac{\mu^2}{v_z^2} = 0$
   in the $|\mu| \gg |m|$ limit.
     This is a simple explanation of the finding
   of \cite{Yamakage12}.  In this respect, thus rigorously
   speaking, RN or AN is not an indication of topological
   character of the underlying normal phase, but rather the
   $\vec d (\vec k)$ configuration of the superconducting phase.

   \section{Conclusion}\label{sec:concl}

   In this paper, we have constructed a pseudospin basis to describe the
   normal state of Bi$_2$Se$_3$.  Superconductivity is then expressed
   in this basis.  Using this approach, many of our previous knowledge
   in unconventional superconductivity can then be directly applied,
   especially for the bulk.  We have also shown that many features of the surface bound
   states can also be understood in this way.
   Although we have concentrated on the surfaces parallel to the Bi$_2$Se$_3$
   quintuple layers in this paper, the same considerations are applicable to other surfaces
   as well as systems with other symmetries.
   This picture however misses the topological properties of the normal phase,
   which in turn some features of the surface bound states in the
   superconducting phases, but only at momenta comparable with or larger than
   the Fermi momenta in the weak-coupling limit.

   \acknowledgements

   This research was supported by
   the National Science Council of Taiwan under grant number
   NSC101-2112-M-001-021-MY3.  I also thank  Lei Hao, T. K. Lee, Liang Fu
   for discussions and communications of their results which
   motivate me to this work, and Bor-Luen Huang for his help in preparing Figure 1.

   \appendix

   \section{ } \label{sec:app}

\begin{figure}
\begin{center}
\rotatebox{0}{\includegraphics*[width=55mm]{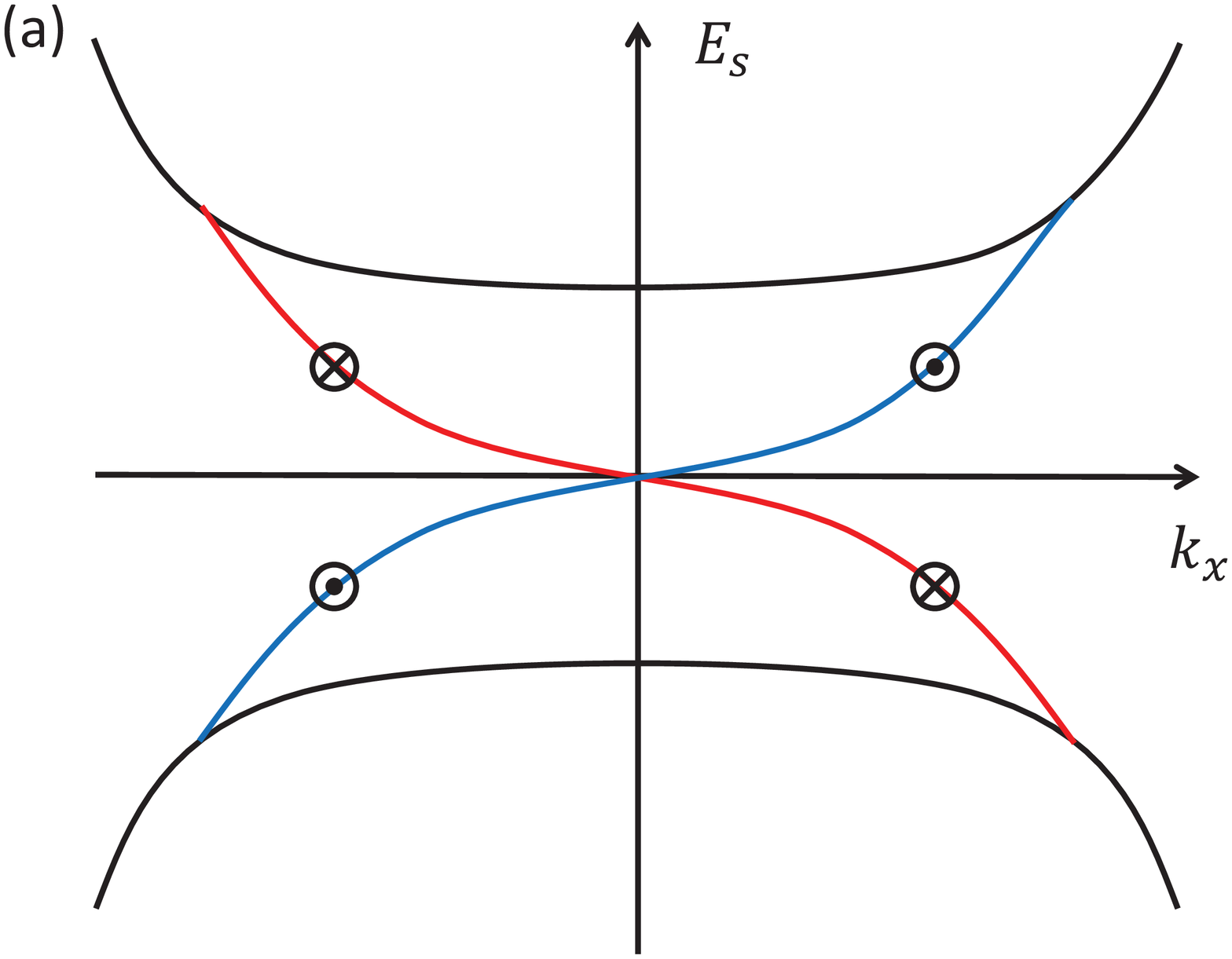}}
\rotatebox{0}{\includegraphics*[width=55mm]{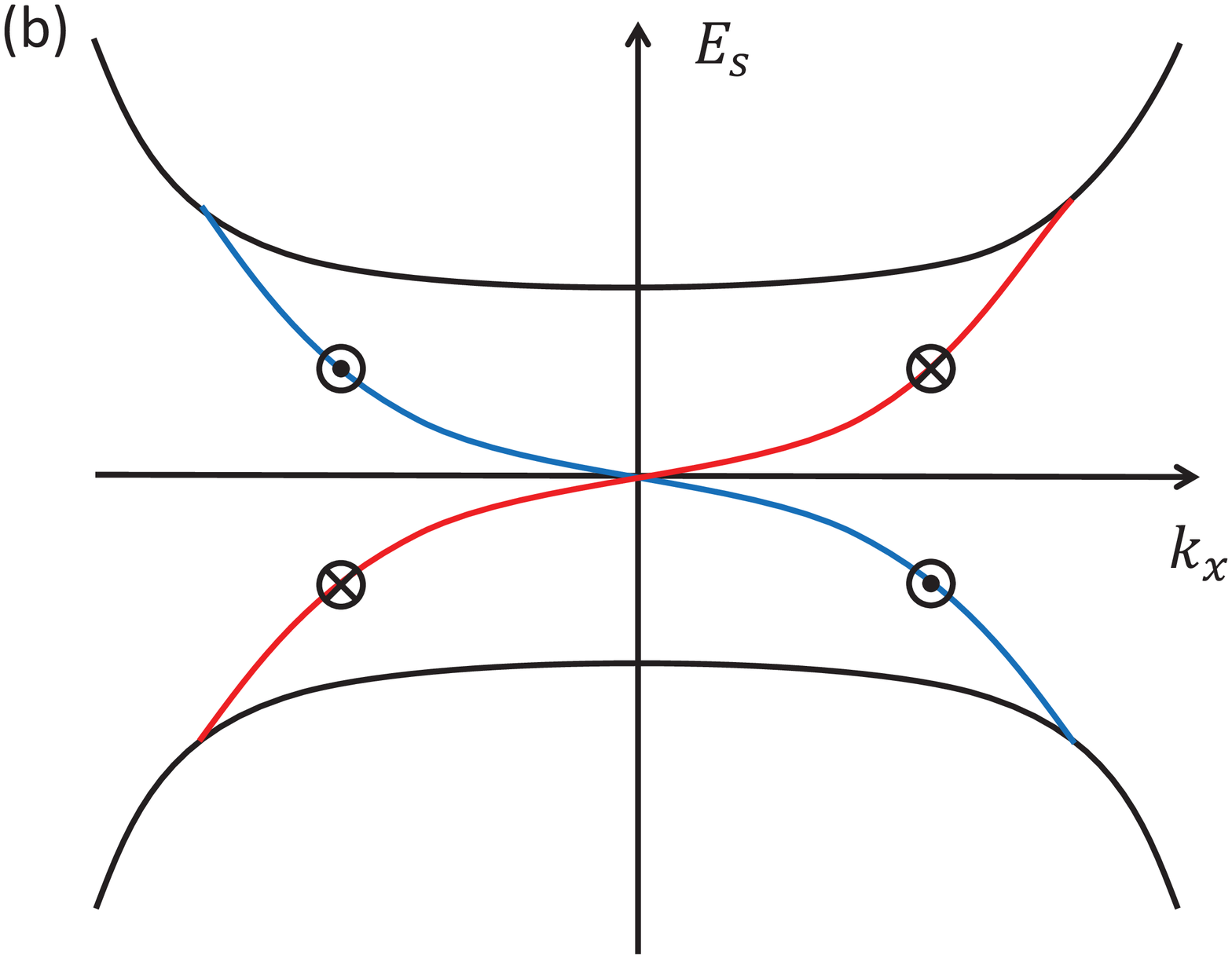}}
\rotatebox{0}{\includegraphics*[width=55mm]{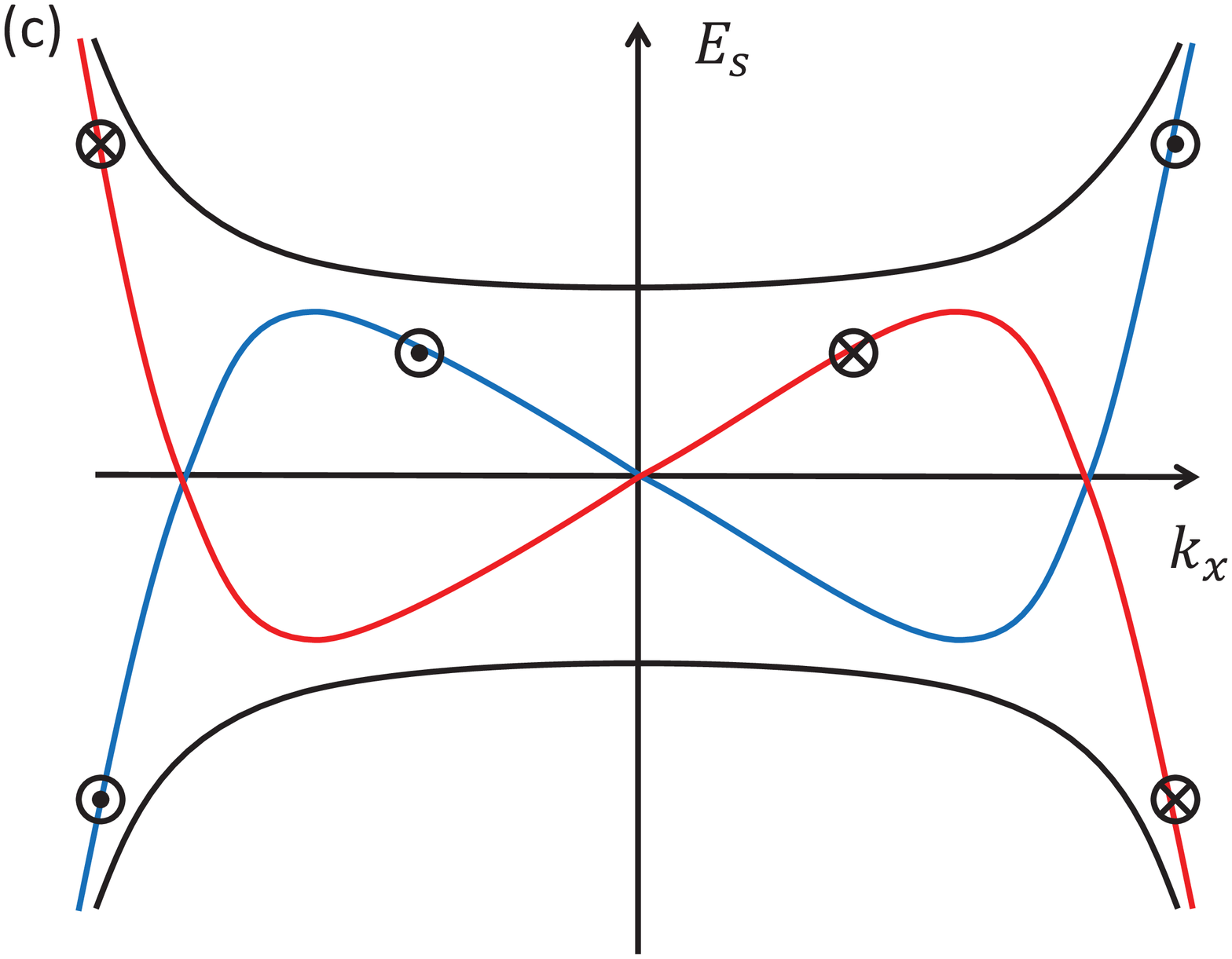}}
\caption{Schematic dispersions for surface bound states in the $A_{1u}$ phase.
(a): Single-band  or
full two-band picture,  ${\rm sgn}(m v_z) > 0$,
(b): Single-band picture, ${\rm sgn}(m v_z) < 0$,
(c): full two-band picture, ${\rm sgn}(m v_z) < 0$.
Here, the arrow head (tail) indicates that the spin is
pointing out of (in to) the plane.  $v < 0$.
}\label{fig1}
\end{center}
\end{figure}

   We consider discuss some topological aspects and the surface
   states for the $A_{1u}$ and $A_{1g}$ phases.  Let us
   first consider the $A_{1u}$ phase, and begin with simple
   continuity arguments.
   For simplicity, we shall consider non-vanishing $v$ and $v_z$,
  varying only $m$.  For a single-band model with $\vec d (\vec k)$
   in eq (\ref{dA1u}) and (\ref{dzA1u}), we note that at $m=0$,
   $\vec d (\vec k)$ only has $\hat z$ component and is odd in $k_z$.
   The bulk state then has a line node on the equator, and
   the surface state is simply a flat band independent of $k_{\|}$.
   This is how the surface state spectra evolve between Fig \ref{fig1}(a)
   and Fig \ref{fig1}(b) when $m$ changes sign.

   Time-reversal symmetric superconductors in three-dimension can
   be characterized by a winding number $W$. \cite{Schnyder08,Sato09,Qi10}  This value can be
   evaluated by first transforming the Hamiltonian into an off-diagonal form.
   The Hamiltonian for the $A_{1u}$ phase in the Nambu-II notation
   is, in the one-band model, $\xi \tau_z + (\vec d (\vec k) \cdot \vec s) \tau_x$.
   Here $\xi$ is the kinetic energy measured with respect to the chemical
   potential $\mu$.  For example, for a quadratic band with particle-like dispersion,
   $\xi = \frac{k^2}{2 M} - \mu$ with $\mu > 0$, whereas for a hole-like band,
    $\xi = - \frac{k^2}{2 M} - \mu$ with $\mu < 0$.  Here $M > 0$ is an effective
    mass (not to be confused with $m$ in eq (\ref{HN})).  The Hamiltonian
    becomes off-diagonal under a rotation in $\vec \tau$ space, such
    as $\tau_z \to \tau_x$, $\tau_x \to \tau_y$.  Then
    the Hamiltonian becomes
    \be
    H_S = \left( \ba{cc} 0 & h_S \\ h_S^{\dagger} & 0 \ea \right)
    \label{Hod}
    \ee
    where $h_S = \xi + i (\vec d (\vec k) \cdot \vec s)$.  The winding
    number can be evaluated from
    \be
    W = \frac{1}{24 \pi^2} \int d^3 {\vec k} \epsilon_{abc}
    {\rm Tr} \left[ q^{\dagger} \frac{ \partial q}{ \partial k_a}
      q^{\dagger} \frac{ \partial q}{ \partial k_b}
      q^{\dagger} \frac{ \partial q}{ \partial k_c} \right]
      \label{W}
      \ee
      where $a,b,c = x, y,z$, $\epsilon_{abc}$ is the fully antisymmetric tensor,
      and $q \equiv h_S / E_S$ is a unitary matrix. Here $E_S = [\xi^2 + \vec d \cdot \vec d]^{1/2}$.
      For the BW phase with $\vec d$ parallel to $\vec k$, we get
      $W = {\rm sgn} \mu$. \cite{amb}  For our state with
      eq (\ref{dA1u}) and (\ref{dzA1u}), we get instead
      \be
      W = ({\rm sgn} \mu)  {\rm sgn} ( \Delta_{1u} v_z)
      \ee
      independent of the sign for $m$, since both $d_{x,y}$ would
      change sign under a sign change of $m$.  Thus, the winding
      number seems insufficient to indicate the possible change
      in the surface state spectra (and the associated bulk topology)
       between Fig \ref{fig1} (a) and (b).

      Now we turn to the full two-band model, and again first employ
      only continuity arguments.  The state is gapped if $\Delta_{1u} \ne 0$
      so long as $\mu^2 + \Delta_{1u}^2 > m^2$.  Hence, one
      can change sign of $m$ without going through any gapless phase, provided
      the above inequality is satisfied.
      Hence when ${\rm sgn} (m v_z)$ changes sign, it cannot affect
      which spin species is connected to the $E_S > 0$ band at large
      $k_{\|}$, even though the sign of the dispersion can change
      for smaller $k_{\|}$.  Hence, continuity argument shows
      that Fig \ref{fig1}(c) should evolve to Fig \ref{fig1}(a)
      when ${\rm sgn} (m v_z)$ changes from $< 0$ to $> 0$.

      We can also examine the winding number $W$ in the full two-band model.
      The Hamiltonian in the Nambu-II notation
      $H_S = (H_N - \mu) \tau_z + \Delta_{1u} \sigma_y s_z \tau_x$ becomes
      off-diagonal by the same rotation in $\vec \tau$ space mentioned
      before, with
      $h_S = ( H_N - \mu) + i \Delta_{1u} \sigma_y s_z$.
        Since the state is gapped so long
      as $\mu^2 + \Delta_{1u}^2 > m^2$ and the winding number
      cannot change within a gapped phase,
       one can first turn off $m$ and then $\mu$, provided $\Delta_{1u} \ne 0$.
       The eigenvalues in eq (\ref{ESA1u}) become degenerate and
       we can use $q = h_S/E_S$, with $q$ unitary, in eq (\ref{W}).
        A direct evaluation gives $W = 1$ (for $\Delta_{1u} > 0$) irrespective of the sign
        of ${\rm sgn} (m v_z)$.  There is no topological phase transition
        between ${\rm sgn} (m v_z) < 0$ and $> 0$ within the $A_{1u}$ phase.
        We note that the single band model therefore produces a spurious
        topological change when $m$ changes sign.

   Lastly, we consider the even parity $A_{1g}$ phase.  In the single-band
   picture, this is the ordinary $s$-wave superconductor.  There is
   no phase difference for the order parameter at $\vec k_{\rm in}$
   and $\vec k_{\rm out}$, and so no bound states are expected.
   The surface states for $A'_{1g}$ and $A^{''}_{1g}$ phases
   have been investigated by \cite{Hao11}.  They showed that
   there are no surface states for $A'_{1g}$, in accordance
   with above.  Interestingly, they found that surface states
   survives for $A^{''}_{1g}$.  The surface states seem to
   be in the form of two Dirac cones, related by the particle-hole
   symmetry of the superconductivity, and crossing each other
   at the chemical potential.  We do not have a simple explanation
   of this in our single-band picture.  Hao and Lee \cite{Hao11}
   noted that the $\Delta^{''}_1$ term does not split the crossing
   at the Fermi level.  It is unclear whether this reflects any topology
   of the $A^{''}_{1g}$ phase, such as a possible even winding
   number that is allowed for a time-reversal symmetric even parity
   superconducting state \cite{Schnyder08}.  We here simply  note
   that, if a general $A_{1g}$ is considered, the pure $A^{''}_{1g}$ state
   and the pure $A'_{1g}$ state are connected in the sense that
   one can find a path in parameter space which connects them
   with the state remaining fully gapped (see Sec \ref{sec:bulk}).
   The $A'_{1g}$ phase is topologically trivial, and no surface
   states are expected.  Thus the surface states of $A^{''}_{1g}$
   are expected to be destroyed in general once a finite $\Delta'_1$ is
   introduced.  This is in accordance with the argument of \cite{Hao11}
   where they showed that $\Delta'_1$ would introduce a finite matrix
   element coupling the two surface states of $A^{''}_{1g}$.

\newpage

\begin{thebibliography}{plain}
\bibitem{BiTe} Bi$_2$Te$_3$ is also found to be superconducting
under pressure \cite{JLZhang11}.  However, the bulk carriers are holes which
are located away from the $\vec k = 0$ $\Gamma$ point, a rather
different situation from Cu:Be$_2$Se$_3$.  We shall not
discuss this compound here any further in this paper.
\bibitem{compare}  Different models have also been employed in the literature, e.g.,
\cite{HZ09,Hao11,Sasaki11}, hence some modifications must be taken to
compare those results with here.  In Model I of \cite{Hao11},
the spin part in eq (\ref{HN}) was instead written as $k_x s_x + k_y s_y$.
Hence all the spins
in \cite{Hao11} must be rotated by $\pi/2$ about z in order to compare
with ours here.  Fortunately, this would not alter anything for
pairings between opposite spins, that is, for the $A_{1g}$, $A_{1u}$ and
$A_{2u}$ phases in Table \ref{table}.  For the two-dimensional odd parity
state $E_u$ which involves pairing between parallel spins,
this rotation brings in the factors $\mp i$ for
$|1 \uparrow 2\uparrow >$ and $| 1 \downarrow 2 \downarrow>$ respectively,
thus interchanges the two entries under $E_u$ in Table \ref{table}.
We shall not try to provide the comparison with other models in this paper.
\bibitem{twoband}  We are using this term with a different meaning
from the usual "two-band superconductivity" in the literature.
Here, we only have
one Fermi surface for the corresponding normal phase
in the absence of pairing potentials.
\bibitem{others}  There are no entries listed under columns
(iii), (iv) and (v) for $A_{2g}$ and $E_g$, since they
do not arise if we insist that there is no momentum dependence
when the pairs are expressed in the orbital and spin basis.
They can however occur if we allow more general momentum dependence.
\bibitem{otherbasis} One is free to work in other basis, e.g. \cite{Michaeli12},
but the physical picture needed would be different.  Also, one would
not be able to employ the tables in \cite{VG84,Ueda85,Blount85,Sigrist91,Yip93,Joynt02}.
\bibitem{notekz0}  We shall not worry about how we define states
at $k_z = 0$.  We can define these by approprate limiting procedures.
This can lead to discontinuities of these wavefunctions as functions
of $\vec k$ (see \cite{Blount85}) , but these complications would not occur for the pair
wavefunctions, our major concern in this paper.
\bibitem{notePT} The pseudospin $\rho_j$ obeys
      $\rho_{j} (\vec k) \to \rho_{j} (- \vec k)$ under parity
      while
       $\rho_{j} (\vec k) \to - \rho_{j} (- \vec k)$ under time-reversal.
       This is applicable to both $\rho'_{1,2,3}$ and $\rho_{x,y,z}$
       defined below due to eq (\ref{P}) and (\ref{T}).
\bibitem{123} We are using $1,2,3$ here for these matrices
in primed basis since they do not have
the rotational properties of $x,y,z$.
\bibitem{phaseshift}  We can separate the reflection due
to the vaccum-sample surface and the Andreev reflection due to the order parameter
variations by imagining inserting a thin normal layer near the surface,
as done in, {\it e.g.} \cite{Beenakker}.  Since the reflection
coefficient for the hole is given by the complex conjugate
of that of the particle, the phase shift $\alpha_{\vec k}$ from the reflection coefficient
$R_{\vec k}$ of Sec \ref{sec:wf} cancels, if we ignore the small
wavevector difference between the particle and hole.
\bibitem{Eb}  These results can be derived easily from solving
the Andreev equation, or locating the poles of the Green's function
in, e.g., \cite{KO}.
\bibitem{amb} There seem to have an ambiguity in the sign of $W$
since one can characterize the same phase with $\vec d $
antiparallel to $\vec k$ by making use of a gauge transformation.
\end{thebibliography}
\end{document}